\documentclass{camera}

\usepackage{amssymb}
\usepackage{amsmath}
\usepackage[dvips]{graphicx}
\usepackage{epsfig}
\usepackage{natbib}

\begin{document}

%
\title{Gamma-ray burst afterglows from trans-relativistic blast wave simulations}

%
\author{H.J. van Eerten$^1$, K. Leventis$^1$, Z. Meliani$^2$ and R.A.M.J. Wijers$^1$
\\ \footnotesize$^1$API, University of Amsterdam, $^2$CfPA, K.U. Leuven}
%
\begin{center}
\LARGE{ \textbf{Gamma-ray burst afterglows from\\ trans-relativistic blast wave simulations}}
\end{center}
\begin{center}
\textsc{H.J. van Eerten$^1$, K. Leventis$^1$, Z. Meliani$^2$ \and R.A.M.J. Wijers
\\ \footnotesize$^1$API, University of Amsterdam, $^2$CfPA, K.U. Leuven}
\end{center}

\section{Introduction}
We present a study of the intermediate regime between ultra-relativistic and nonrelativistic flow for gamma-ray burst afterglows. The hydrodynamics of spherically symmetric blast waves is numerically calculated using the \textsc{amrvac} adaptive mesh refinement code. Spectra and light curves are calculated using a separate radiation code that, for the first time, links a parametrisation of the microphysics of shock acceleration, synchrotron self-absorption and electron cooling to a high-performance hydrodynamics simulation.

All results are presented in more detail in \cite{vanEerten2009c}.

\section{Dynamics}
We used typical physical parameters: explosion energy $10^{52}$ erg, circumburst number density 1 cm$^{-3}$, assuming a homogeneous environment, shocked magnetic field energy density a fraction 0.01 of thermal energy density, shocked electron energy density a faction 0.1 of thermal energy density. We trace the evolution of the magnetic field  and gradually decrease the fraction $\xi_N$ of the number density for accelerated particles at the shock from 1 downto 0.1. 

The blast wave slows down slower than anticipated from equating the explosion energy to the total swept-up rest mass energy, which in this case would lead to a transition time $t_{\mathrm{NR}} \backsim 450$ days. We find $\backsim 1290$ days in the emission time frame. The analytical estimate for blast wave velocity $\times$ Lorentz factor ($\beta \gamma$) from \cite{Huang1999} overestimates the late time $\beta \gamma$ by a factor $4/3$.

The simulation uses an advanced equation of state that resulting in an effective adiabatic index $\Gamma_\mathrm{ad}$ gradually evolving from 4/3 to 5/3 from relativistic to nonrelativistic. We find that the lab frame density directly behind the shock $D$, divided by $\rho_0 \gamma^2$ (where $\rho_0$ the circumburst density) remains numerically close to four \emph{throughout the entire simulation}:
$ D / \rho_0 \gamma^2 = ( \Gamma_\mathrm{ad} + 1 / \gamma ) / ( \Gamma_\mathrm{ad} - 1 ) \backsim 4.$
 
\section{Radiation and GRB030329}
\begin{figure}
\includegraphics[width=0.49\columnwidth]{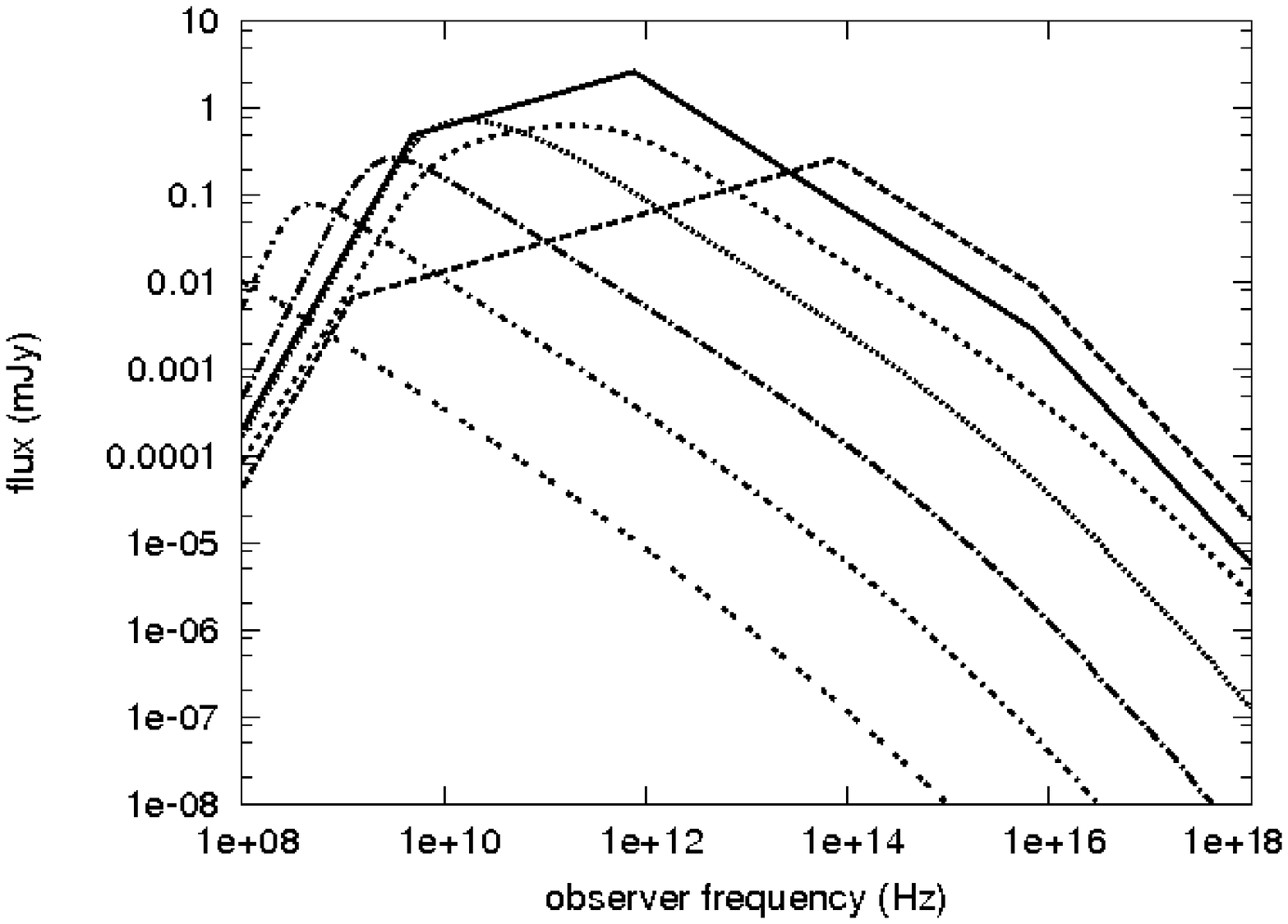}
\includegraphics[width=0.49\columnwidth]{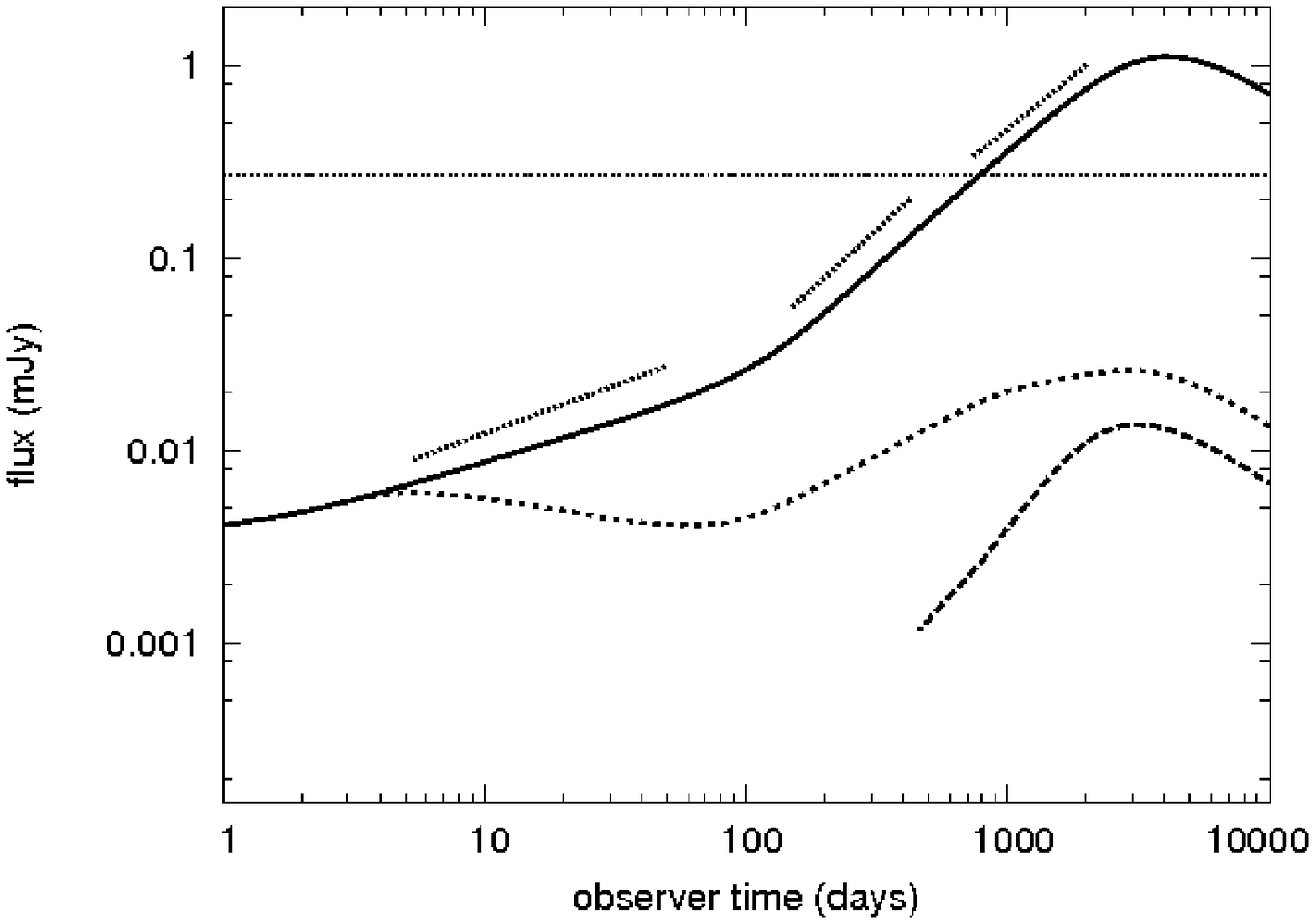}
\caption{\footnotesize Left: Spectra at different observer times. The smooth curves show simulated spectra at different observer times: 1, 10, 100, 1,000 and 10,000 days, with later observed spectra having lower flux in the high frequency range. For comparison we have included predicted slopes at the different power law regimes after 1 day, for both $\xi_N = 1$ (solid line) and $\xi_N = 0.1$ (dashed line). Right: Simulated light curve at 200 MHz for GRB030329, top curve for spherical explosion and bottom curve for hard-edged jet with opening angle 22 degrees. We have drawn the following slopes from left to right: $1/2$, $5/4$ and $11/10$. LOFAR sensitivity for 25 core and 25 remote stations after four hours of integration time is 0.273 mJy and indicated by the horizontal line. The contribution from the counterjet is also shown separately.}
\label{figure} 
\end{figure}
Our code successfully reproduces the synchrotron spectrum, including self-absorption and electron cooling (see fig. \ref{figure}). The transition to the nonrelativistic regime occurs around $~1000$ days in observer time. Only the relativistic slopes of the light curve match the analytically expected values. The changing $\Gamma_\mathrm{ad}$ leads to a late-time steepening, but the light curve is then more strongly influenced by the change in $\xi_N$, leading to a less steep decay.

Simulating GRB030329 using explosion parameters derived from analytical model fits reveals a strong discrepancy between simplified model and more accurate simulation, with flux differences up to an order of magnitude. Simulated late time radio light curves using hard-edged jets indicate that the counterjet should eventually become visible, and show a doubly peaked curve due to the combination of strong self-absorption and the jet break.
\footnotesize
\bibliographystyle{bst/aa}
\bibliography{hveerten}

\begin{thebibliography}{2}
\expandafter\ifx\csname natexlab\endcsname\relax\def\natexlab#1{#1}\fi

\bibitem[{{Huang} {et~al.}(1999){Huang}, {Dai}, \& {Lu}}]{Huang1999}
{Huang}, Y.~F., {Dai}, Z.~G., \& {Lu}, T. 1999, \mnras, 309, 513

\bibitem[{{Van Eerten} {et~al.}(2009){Van Eerten}, {Leventis}, {Meliani},
  {Wijers}, \& {Keppens}}]{vanEerten2009c}
{Van Eerten}, H.~J., {Leventis}, K., {Meliani}, Z., {Wijers}, R.~A.~M.~J., \&
  {Keppens}, R. 2009, MNRAS accepted. ArXiv: 0909.2446

\end{thebibliography}

\end{document}